# Modelstamp: Pre-Deserialization Verification of Machine-Learning Artifacts and Runtime Environment State

Anagha Dhekne
*Independent Researcher*

Abstract—Persisted machine-learning models can remain byte-identical while the software environments in which they are loaded evolve, creating a verification problem that artifact integrity checks alone cannot expose. This paper presents Modelstamp, a lightweight Python persistence library for verifying artifact integrity and represented runtime-environment state before deserialization. At persistence time, Modelstamp associates a serialized artifact with a sidecar JSON manifest containing a SHA-256 digest, runtime metadata, and installed versions from a bounded tracked-package set; a separately recorded model-relevant subset determines which package versions participate in drift comparison. Optional HMAC authentication supports workflows in which the producer and verifier share a secret key. At verification time, the artifact and represented current environment are checked against this recorded evidence before the model is deserialized. Modelstamp is evaluated using 14 controlled environment-drift scenarios, eight controlled trust-boundary scenarios, and an artifact-size scaling benchmark from 10 MiB to 1 GiB. The controlled drift experiments behaved as specified across relevant dependency changes, unchanged environments, and unrelated environmental changes, including broader noise controls. The trust-boundary experiments similarly confirmed both intended detections and expected limitations, including shared-key forgery and replay. Median verification time increased from 0.032 s at 10 MiB to 3.334 s at 1 GiB, with measured throughput of approximately 307–312 MiB/s in the benchmark environment. These results characterize Modelstamp as a complementary pre-deserialization reference-state verification control rather than as a replacement for dependency-management systems, malicious-model detection, safe deserialization, or public publisher authentication.

## I. INTRODUCTION

Machine-learning models are routinely serialized so that they can be stored, transferred, deployed, and reused without retraining. In Python-based workflows, however, a persisted model is not necessarily independent of the software environment in which it is later consumed. Its successful loading and behavior may depend on the Python runtime, machine-learning framework, and supporting libraries present at deserialization time. A model artifact can therefore remain byte-identical while the environment around it evolves, creating a condition that is invisible when only the serialized file itself is considered.

This creates two distinct verification concerns. The first is artifact integrity: whether the serialized bytes being loaded are the same bytes that were previously established as the reference artifact. The second is environment drift: whether model-relevant runtime dependencies at verification time differ from those recorded when the artifact was persisted. Cryptographic hashing can identify byte-level modification of an artifact, but an unchanged digest cannot reveal that a relevant dependency has changed. Conversely, a dependency-version difference does not imply that the artifact itself has been modified, nor does it necessarily mean that model behavior has changed. Artifact integrity and environment consistency should therefore be treated as related but separate properties of persisted-model verification.

Existing research and tooling address important aspects of ML artifact trust and reuse, including unsafe serialization, malicious-artifact detection, constrained deserialization, publisher authentication, dependency representation, environment capture, and software-artifact verifiability. Model-persistence and lifecycle tooling can also record dependency metadata or warn about cross-version incompatibility. These mechanisms answer different questions about an artifact's origin, contents, execution, reproducibility, or software context. The concern investigated here is narrower: given an ML artifact whose reference state has already been established, can artifact consistency and a model-oriented projection of recorded runtime state be checked before deserialization while keeping the limits of those checks explicit?

This paper presents Modelstamp, a lightweight Python persistence library designed to investigate this problem. When an ML artifact is persisted, Modelstamp creates a companion manifest containing artifact-integrity information and runtime metadata. The current implementation records Python/runtime fields and installed versions from a bounded tracked-package set, while a separately recorded model-relevant subset determines which package versions participate in drift comparison. The artifact is associated with a SHA-256 digest, and an optional HMAC mechanism can authenticate the recorded evidence in controlled shared-secret workflows. During subsequent verification, Modelstamp checks the artifact against its recorded integrity evidence and compares the represented current environment with the applicable recorded state before deserialization occurs.

A central design consideration is determining which recorded package changes are relevant to the persisted model. Comparing every installed Python package can generate warnings for changes that have no relationship to the model being loaded. Modelstamp therefore separates bounded package-state capture from model-specific comparison: the manifest records installed versions from the tracked set, and relevance rules select a subset for package-drift evaluation. This introduces an empirical question: whether relevant dependency changes can be surfaced without producing warnings for unrelated environmental changes. It also creates a clear limitation: the selected subset is an approximation of model relevance rather than a complete representation of every factor capable of affecting execution.

The evaluation examines Modelstamp along three dimensions. First, a 14-scenario controlled environment-drift validation examines framework and supporting-dependency changes alongside identical-environment and unrelated-package controls. Second, an eight-scenario controlled trust-boundary validation examines artifact integrity and optional authentication under both detectable modifications and conditions intentionally outside the mechanism's guarantees. Third, an artifact-size scaling benchmark measures verification cost for artifacts ranging from 10 MiB to 1 GiB, using repeated measurements with verification isolated from deserialization. These experiments address the following research questions:

RQ1 — Environment drift: In controlled persisted-model scenarios, can Modelstamp distinguish changes in model-relevant runtime dependencies from unrelated environmental changes?

RQ2 — Trust boundaries: Which artifact-integrity and authentication threats can Modelstamp detect before deserialization, and which remain possible under its stated trust model?

RQ3 — Performance: How does the computational cost of Modelstamp verification scale across persisted artifacts ranging from 10 MiB to 1 GiB?

This paper makes three contributions. First, it investigates a lightweight artifact-bound persistence design that combines cryptographic artifact-integrity evidence, bounded runtime-state capture, model-relevance-filtered package comparison, and pre-deserialization verification under an explicit trust model. This combination is positioned as complementary to existing dependency-management, persistence, and artifact-security mechanisms rather than as the invention of dependency recording or cross-version warnings. Second, the evaluation shows that Modelstamp produced the expected distinction between relevant and unrelated environmental changes across all 14 controlled scenarios, including broader environmental-noise cases in which multiple unrelated package versions changed while relevant drift remained independently detectable. Third, the experiments characterize both the limits and operational cost of the verification model: all eight trust-boundary scenarios behaved according to the stated guarantees, including cases intentionally accepted because hashing or shared-secret authentication cannot provide provenance or freshness guarantees; across artifacts ranging from 10 MiB to 1 GiB, median verification time increased from approximately 0.032 s to 3.334 s, while measured throughput remained approximately 307–312 MiB/s in the benchmark environment.

Modelstamp is not intended to establish that an artifact is intrinsically benign or correct, provide public publisher identity, or constrain code execution during deserialization. Rather, it complements mechanisms addressing those concerns by making artifact identity and recorded model-relevant environment state explicit and verifiable before deserialization.

## II. BACKGROUND AND PROBLEM DEFINITION

### *A. Persisted Machine-Learning Artifacts and Model Reuse*

Machine-learning workflows commonly separate model training from later inference and deployment. After training, a model can be serialized into a persistent artifact and subsequently reconstructed without repeating the training process. This enables model reuse across processes, systems, and deployment environments, but it also means that the persisted artifact becomes part of the software supply chain through which trained models are distributed and consumed. Empirical work on pretrained-model reuse has shown that practitioners consider attributes such as provenance, reproducibility, and portability when evaluating reusable models, while also reporting missing metadata, discrepancies between reported and observed performance, and security concerns as barriers to trustworthy reuse [1].

For Python-based machine learning, persistence frequently relies on serialization mechanisms such as pickle, joblib, and framework-specific formats built on related serialization mechanisms. These formats can encode substantially more than numerical model parameters. In particular, unsafe Python serialization mechanisms can cause executable behavior to occur during object reconstruction. Large-scale empirical analysis of publicly distributed ML models has demonstrated that unsafe serialization formats remain common and that malicious payloads can be triggered during deserialization [2]. This makes the transition from a stored artifact to a reconstructed Python object a security-sensitive boundary.

The present work does not attempt to determine whether serialized content is malicious or to make an unsafe serialization mechanism safe. Instead, the significance of this boundary is that checks concerning the identity and recorded runtime state of an artifact should occur before deserialization, when possible. Once deserialization of an unsafe artifact has begun, verification performed afterward cannot serve as a pre-load control against discrepancies that could have been identified beforehand.

A persisted model should also not necessarily be treated as self-contained. Reconstructing and using it may depend on the Python runtime, the ML framework under which it was created, and supporting libraries used by the estimator or its components. Consequently, two forms of state are relevant to the problem considered here: the serialized artifact itself and the software environment associated with that artifact. Either can differ between persistence and subsequent verification.

### *B. Artifact Integrity and Authentication*

Artifact integrity concerns whether a retrieved artifact remains byte-consistent with an established reference. Let A denote the reference artifact and H a cryptographic hash function. Its recorded digest is

$$d = H(A) \quad (1)$$

For a candidate artifact A′, the digest is recomputed as

$$d' = H(A') \quad (2)$$

A byte-level discrepancy is detected when

$$d' \neq d. \quad (3)$$

Integrity relative to a digest is not equivalent to authentication, however. If an adversary can replace both an artifact and an unauthenticated digest associated with it, the replacement pair can remain internally consistent. A successful hash comparison consequently answers whether the current artifact bytes correspond to the recorded digest, but does not by itself establish who created the artifact or the digest.

A keyed message authentication code adds a shared secret K. For authenticated manifest evidence M, Modelstamp conceptually computes

$$t = HMAC_K(M) \quad (4)$$

where t is the authentication tag. A verifier possessing the same secret can test whether the evidence is consistent with a tag generated by a holder of that secret.

Shared-secret authentication still differs from public publisher authentication. Every legitimate holder of the secret is capable of producing valid authentication evidence, so HMAC does not establish non-repudiable publisher identity. Prior work on ML supply-chain security has separately explored Sigstore-based model signing, illustrating the distinct problem of binding model artifacts to verifiable publisher identity [3]. The training-data verification component proposed in that work addresses a separate provenance problem and is outside the scope considered here.

These distinctions motivate treating byte-level integrity, shared-secret authentication, and public publisher identity as separate security properties rather than using “artifact trust” as an undifferentiated concept.

### *C. Runtime Environments and Dependency Drift*

The runtime environment associated with a persisted model can evolve independently of the artifact. Package upgrades, dependency resolution, environment recreation, or deployment to another system can result in different Python or library versions even when the serialized model bytes remain unchanged.

This presents a second verification problem. Suppose a model is persisted while a particular set of runtime packages is installed and is later consumed after one or more of those packages have changed. An artifact-integrity check can still succeed because no artifact bytes have been modified. The changed software state is therefore not observable from the artifact digest alone.

One possible approach to describing software state is to reconstruct dependencies from project declarations, lockfiles, or other package metadata. However, prior research on Python SBOM generation has demonstrated that such reconstruction can be incomplete or inaccurate [4]. Differences among dependency-declaration mechanisms, lockfile formats, version specifications, and tool implementations can cause dependencies or exact versions to be omitted or incorrectly represented. This evidence concerns dependency reconstruction from project metadata, rather than persisted ML artifacts specifically.

The problem considered here takes a different approach to environment evidence. Instead of attempting to infer the persistence-time environment later from static dependency declarations, concrete installed package versions can be observed at persistence time and recorded as part of the reference state. This avoids the specific reconstruction problem documented for declaration-based tooling: the recorded value represents installed state observed at that moment rather than an inference about what a declaration should resolve to.

Directly observing installed packages does not, however, solve the entire problem. A Python environment may contain many packages unrelated to a particular model. Treating every installed package as equally relevant can cause unrelated upgrades to appear as meaningful model-environment drift. The remaining challenge is therefore dependency relevance: determining which portion of the observed environment should be associated with a particular persisted artifact.

This work uses environment drift to mean a difference between recorded and current state for a dependency identified as model-relevant. The term does not imply that the change is harmful or incompatible. A version change may have no observable effect on a model, while behaviorally significant differences may also arise from runtime factors not captured by package-version metadata. Environment drift is therefore evidence of a recorded state difference, not evidence of behavioral failure.

### *D. Reproducibility and Verifiability*

Environment verification is related to, but distinct from, software reproducibility. Reproducibility generally concerns whether an artifact or result can be recreated from specified inputs, dependencies, and build or execution conditions. Verification against a reference instead asks whether a candidate state remains consistent with evidence established for a previously known state.

Prior work on decentralized software-package ecosystems has explicitly distinguished reproducible builds from independently verifiable artifacts and has shown that dependency and build-environment recovery can affect the ability to establish these properties [5]. That work concerns general software-package ecosystems—including systems such as npm, PyPI, crates.io, and RubyGems—rather than ML model artifacts specifically. Its relevance here is therefore conceptual: reproducibility and verifiability should not be treated as interchangeable properties.

For persisted ML models, recording environment state does not make the model reproducible in the stronger sense of reconstructing its training process, training data, hardware configuration, or complete software stack. Likewise, verifying that recorded package versions still match does not establish that the model could be retrained to produce an equivalent artifact. The narrower property considered in this paper is whether an already-established artifact and selected information about its associated runtime environment remain consistent at a later verification point.

### *E. Problem Definition*

The preceding concepts make precise the terms reference state, artifact consistency, and environment drift used throughout this paper.

Let A denote a serialized machine-learning artifact established as part of a trusted reference state. Let T denote the bounded package set tracked by the implementation, and let R(A) denote the model-relevant subset selected for artifact A, such that

$$R(A) \subseteq T. \quad (5)$$

The package state captured at persistence time is

$$E_s\text{^}cap = \{(p_i, v_i) \mid p_i \in T \text{ and } p_i \text{ is installed}\}. \quad (6)$$

Package comparison uses the relevance-filtered projections

$$E_s\text{^}rel(A) = \{(p_i, v_i) \in E_s\text{^}cap \mid p_i \in R(A)\}, \quad E_l\text{^}rel(A) = \{(p_i, v'_i) \mid p_i \in R(A) \text{ and } p_i \text{ is currently installed}\}. \quad (7)$$

Modelstamp also records and compares runtime-level state independently of package relevance. Let the recorded and current runtime tuples be

$$E_s\text{^}rt = (py_s, impl_s, plat_s), \quad E_l\text{^}rt = (py_l, impl_l, plat_l). \quad (8)$$

For verification, the represented recorded and current environment states are therefore

$$E_s = (E_s\text{^}rel(A), E_s\text{^}rt), \quad E_l = (E_l\text{^}rel(A), E_l\text{^}rt). \quad (9)$$

Package-version drift is the set-valued function

$$D_pkg(E_s, E_l; A) = \{p_i \in R(A) \mid v_i \neq v'_i \text{ or } p_i \text{ is absent from the current relevant state}\}. \quad (10)$$

The broader environment-drift predicate is Boolean and is true when either the package drift set is nonempty or the runtime tuple differs:

$$D(E_s, E_l; A) \equiv [D_pkg(E_s, E_l; A) \neq \emptyset] \vee [E_s\text{^}rt \neq E_l\text{^}rt]. \quad (11)$$

This definition distinguishes what is captured from what is compared: E_s^cap preserves installed versions from the bounded tracked set, R(A) selects the package projection used for model-specific comparison, and runtime-level fields are compared independently of package relevance.

Artifact consistency is evaluated separately. Given recorded digest d, consistency requires

$$H(A') = d, \quad (12)$$

whereas an artifact-integrity discrepancy is present when

$$H(A') \neq d. \quad (13)$$

Artifact consistency and recorded-environment consistency are intentionally independent. This yields four conceptual states:

**TABLE I**
**FOUR CONCEPTUAL VERIFICATION STATES**

| Artifact | Env. | Interpretation |
|---|---|---|
| Match | Match | No discrepancy relative to the recorded reference state. |
| Mismatch | Match | Artifact-integrity discrepancy. |
| Match | Drift | Environment changed while the artifact remained byte-consistent. |
| Mismatch | Drift | Artifact and recorded environment both differ from the reference state. |

The broader reference state encompasses the artifact and evidence established about it at persistence time, including recorded runtime information and, where applicable, integrity and authentication evidence. The notation above deliberately formalizes individual verification properties rather than reducing the entire reference state to a pair such as S = (A, E), which would omit security evidence associated with that state.

The problem addressed in this paper can therefore be stated as follows: given an established reference artifact and recorded runtime evidence, determine before deserialization whether a candidate artifact remains consistent with its integrity evidence and whether the represented environment state has drifted under D(E_s, E_l; A). Authentication may additionally establish consistency with evidence generated by a holder of an authorized shared secret, subject to the trust assumptions defined in Section III.

This formulation does not ask whether the candidate model is benign, behaviorally compatible, reproducible, or safe to deserialize. Those properties require different evidence and mechanisms. Instead, it defines the narrower verification problem addressed by Modelstamp and provides the conceptual basis for the threat model and scope that follow.

## III. THREAT MODEL AND SCOPE

Modelstamp operates under a trusted-reference-state assumption. The model artifact and its associated metadata are assumed to be trustworthy at the time the reference state is established. Modelstamp does not determine whether the original model is benign, correctly trained, or obtained from a trustworthy publisher. Instead, it establishes evidence about that reference state so that subsequent changes to the artifact or its recorded model-relevant environment can be identified before deserialization. Consequently, an artifact that is already malicious when first persisted may still pass later Modelstamp verification if it remains consistent with the recorded reference state.

### *A. Protected Assets and Verification Boundary*

Modelstamp considers two primary persisted components: the serialized model artifact and its associated sidecar JSON manifest. The manifest records the artifact's SHA-256 digest together with runtime metadata used for environment comparison, including the Python version and selected model-relevant package versions. When authentication is enabled, HMAC evidence is additionally associated with the protected state.

The relevant verification boundary lies between retrieval of the persisted artifact and its deserialization. Modelstamp performs applicable authentication, artifact-integrity, and environment-state checks before the serialization library is allowed to reconstruct the model object. This ordering limits the mechanism to pre-deserialization verification; it does not constrain what an artifact may execute if deserialization subsequently proceeds.

### *B. Adversary Capabilities*

The threat model considers an adversary who may obtain access to persisted model files or their associated manifests after the trusted reference state has been established. Depending on the scenario, such an adversary may modify the serialized artifact while leaving the manifest unchanged; modify integrity or descriptive information stored in the manifest; replace an artifact and its manifest together; attempt to replace authenticated evidence with unsigned evidence; provide evidence authenticated using a key that is not trusted by the verifier; or replay an older artifact-and-manifest pair that was previously valid.

The threat model also considers the stronger case in which a party possesses a legitimate shared HMAC secret. Such a party can construct new evidence that is cryptographically valid under that secret. This case distinguishes the properties of shared-secret authentication from stronger guarantees such as independent publisher identity or non-repudiation.

Each of these adversarial capabilities is exercised by one or more controlled scenarios in the RQ2 trust-boundary evaluation, allowing observed implementation behavior to be compared directly with the guarantees defined here.

The threat model does not assume that Modelstamp controls the host operating system, Python interpreter, filesystem, or process executing verification. An adversary with sufficient control to modify Modelstamp itself, replace the verification code, extract trusted secrets, or alter verification results is outside the protection boundary considered in this study.

### *C. Integrity and Authentication Properties*

For artifact integrity, Modelstamp records a SHA-256 digest of the persisted artifact. At verification time, the digest is recomputed from the candidate artifact and compared with the recorded value. A mismatch indicates that the candidate artifact differs at the byte level from the artifact represented by the recorded digest. This establishes consistency relative to the recorded digest; by itself, it does not establish who created either the artifact or the digest.

Optional HMAC authentication addresses a different property. In workflows where the producer and verifier share a trusted secret, HMAC can provide evidence that the authenticated state was generated by a party possessing that secret and has not subsequently been altered without recomputing valid authentication evidence. However, possession of the shared secret also grants the ability to generate new valid evidence. Modelstamp's HMAC mechanism therefore does not provide public publisher identity, non-repudiation, or protection against forgery by an authorized key holder.

Neither SHA-256 nor HMAC inherently establishes freshness. A previously valid authenticated artifact-and-manifest pair may remain cryptographically valid if replayed later. Preventing rollback or replay would require an additional freshness mechanism, such as trusted version state, monotonic counters, timestamps backed by a trusted authority, or an external transparency mechanism. Such mechanisms are outside the current design.

These properties are deliberately separated. SHA-256 provides evidence about byte-level artifact integrity, while HMAC can additionally authenticate protected evidence within a shared-secret trust relationship. Neither property alone establishes public provenance or freshness.

### *D. Environment-Drift Boundary*

Modelstamp considers changes to the runtime environment independently of artifact modification. At persistence time, selected model-relevant package information and runtime-level information are recorded in the manifest. At verification time, corresponding information from the current environment is compared with the recorded state. A relevant package-version difference, or a difference in the recorded Python version, Python implementation, or platform, is reported as environment drift, consistent with the definition in Section II.

Environment drift should not be interpreted as proof of incompatibility. A changed package version may remain fully compatible with the persisted model, while behaviorally significant changes may arise from factors that are not represented in the manifest. Modelstamp therefore provides evidence that recorded environment state has changed; it does not predict whether that change will alter model outputs or cause deserialization to fail.

The environment representation is intentionally selective rather than a complete snapshot of the execution platform. The current design focuses on Python/runtime fields and packages identified as relevant to the persisted model. Factors such as operating-system libraries beyond the recorded platform string, hardware, GPU drivers, CUDA versions, external services, and other native runtime dependencies are not comprehensively represented. Likewise, dependency-relevance discovery is an approximation and may not identify every package capable of influencing an arbitrary model.

Accordingly, absence of detected drift means that the recorded state is consistent with the current state for the information being compared. It does not establish equivalence of the complete execution environment.

### *E. Scope and Non-Goals*

Modelstamp is intended to provide a pre-deserialization verification layer for an already-established reference artifact and its recorded model-relevant environment. Several related security, availability, and ML-assurance problems are deliberately outside this scope.

In particular, Modelstamp does not determine whether the reference model contains malicious code; make pickle, joblib, or other unsafe serialization formats safe to deserialize; sandbox or restrict code execution during deserialization; detect poisoned training data, backdoored weights, or adversarial model behavior; establish the identity of a public model publisher; provide asymmetric signatures, non-repudiation, or a transparency log; guarantee freshness or prevent replay of previously valid evidence; prevent denial of service through deletion, corruption that makes files unavailable, or withholding of the artifact or manifest; guarantee behavioral compatibility when dependency versions match or differ; provide a complete reconstruction of the original execution environment; or establish model correctness, predictive performance, fairness, or regulatory compliance.

These exclusions distinguish Modelstamp from mechanisms concerned with malicious-artifact detection, constrained deserialization, public signing infrastructure, availability, and complete reproducibility. Such mechanisms may be used alongside Modelstamp because they address properties outside its verification boundary.

### *F. Trust Assumptions*

The guarantees evaluated in this work depend on four principal assumptions: (1) the reference state is trustworthy when established; Modelstamp preserves evidence about an established state but does not establish the intrinsic trustworthiness of that state; (2) the verification implementation and execution process are trusted, and compromise of the verifier itself is outside the evaluated protection boundary; (3) HMAC secrets, when used, are protected from unauthorized parties, because any holder of a legitimate shared secret can generate valid authentication evidence; and (4) the recorded environment is a selective representation, so successful environment verification means consistency with the recorded information rather than equivalence of the complete execution environment.

These assumptions define the interpretation of the trust-boundary experiments evaluated under RQ2. In particular, scenarios such as shared-key forgery and replay are expected limitations rather than verification failures: they exercise security properties that SHA-256 and shared-secret authentication are not designed to provide.

This threat model therefore establishes the boundary within which Modelstamp's verification results should be interpreted. The following section describes how the persistence, manifest, dependency-relevance, integrity, authentication, and pre-deserialization verification mechanisms implement these properties.

## IV. MODELSTAMP DESIGN

Modelstamp implements the verification problem defined in Section II-E as a persistence layer around Python model serialization. Its design associates a serialized artifact with a sidecar JSON manifest containing artifact evidence, runtime information, model metadata, a bounded tracked-package snapshot, and a selected model-relevant package subset used for comparison. Verification is performed against this recorded evidence before deserialization. Optional HMAC authentication extends the design for workflows in which the producer and verifier share a trusted secret. Fig. 1 summarizes the persistence and verification paths.

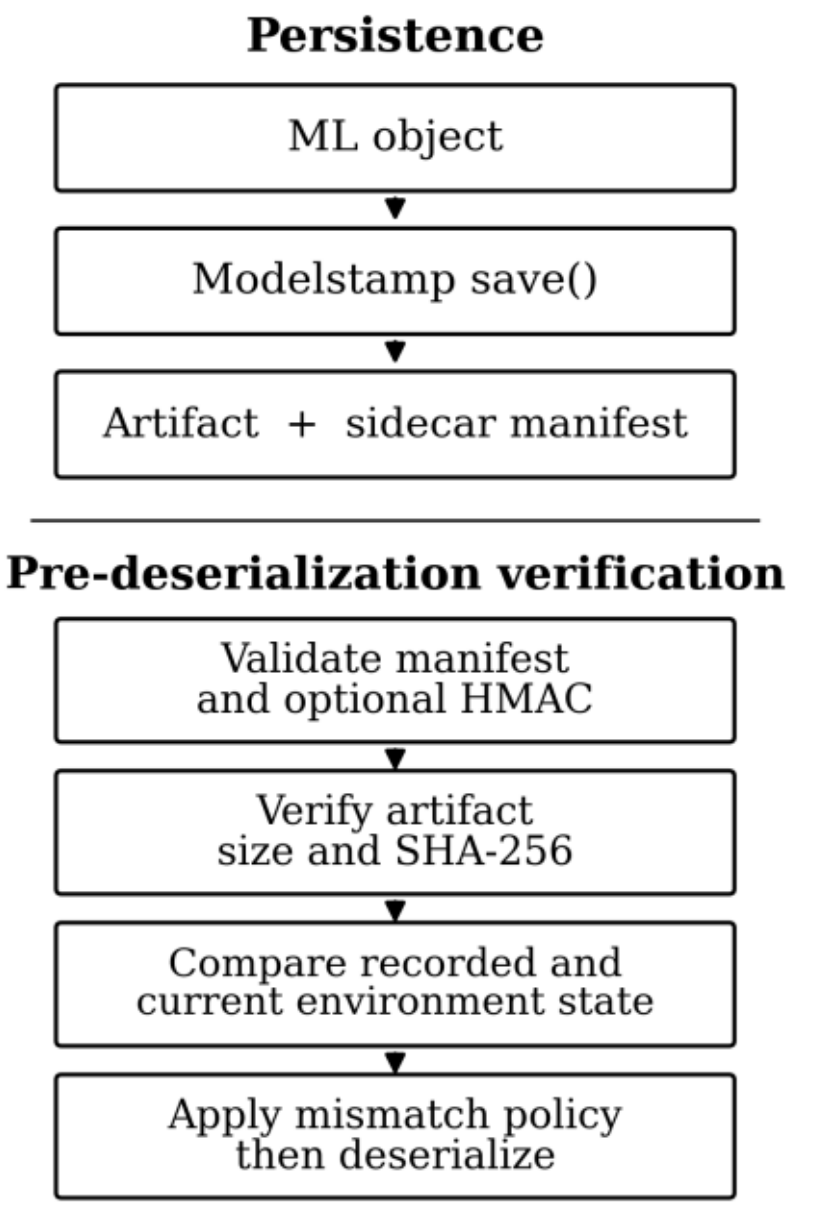

Fig. 1. Modelstamp persistence and pre-deserialization verification workflow. The sidecar manifest records artifact evidence and selected runtime state; verification precedes deserialization.

*A. Design Goals*

The design is guided by five goals. First, verification should precede deserialization. Artifact-integrity and applicable authentication checks should occur before pickle or joblib reconstructs the persisted object. Second, artifact integrity and environment drift should remain separate properties. A model can remain byte-consistent while its runtime environment changes, and a modified artifact can occur without a relevant dependency change. Third, environment comparison should focus on model-relevant dependencies rather than treating every package difference as equally meaningful. Fourth, authentication should remain optional and its shared-secret trust model explicit. Finally, the persistence interface should remain lightweight and should not require a model registry, external verification service, or modification to the underlying serialized model format.

*B. Persistence Workflow*

The persistence workflow begins with a Python object and destination artifact path. The serialization backend may be selected explicitly as pickle or joblib; in automatic mode, joblib is selected when available and pickle otherwise. Modelstamp first serializes the object to a temporary file in the destination directory, then derives model information and the relevant-package set, captures runtime information, computes the artifact's SHA-256 digest and size, and constructs the manifest. Optional HMAC authentication is then applied when a signing key is supplied.

The artifact and manifest are staged before replacement of the destination pair. Modelstamp also serializes operations on the same artifact using thread- and process-level locking. Because two filesystem paths cannot generally be replaced as a single filesystem transaction, the implementation minimizes the interval between replacements and contains rollback logic intended to restore the previous artifact if committing the new pair fails. This provides stronger failure handling than independently writing the artifact and manifest, while not claiming a true two-file atomic filesystem transaction.

The resulting sidecar is named by appending .manifest.json to the artifact filename. For example, model.joblib is associated with model.joblib.manifest.json.

*C. Manifest Structure*

The current manifest schema has version 1 and contains the artifact filename, SHA-256 digest, and size; serialization backend; model class, module, and supported component information; environment information; selected relevant packages; optional user metadata; and optional HMAC-SHA256 authentication information.

The environment object records Python version, Python implementation, platform, package versions, and creation timestamp. When Git capture is enabled and a usable repository is available, the current commit and worktree dirty state are also recorded. The serialization backend is retained so that automatic loading can use the same backend with which the artifact was persisted. Manifest parsing validates required field types and supported values before the recorded information is used.

*D. Model-Relevant Dependency Selection*

The relevance mechanism implements the R(A) concept introduced in Section II using the model's Python module and, for supported composite structures, component modules. For the top-level object, Modelstamp records its class and module and extracts the root module name. When the object exposes a valid steps structure, such as a scikit-learn pipeline, the class and module of each step are additionally recorded and their root modules participate in relevance discovery.

The current implementation maps recognized modules to predefined dependency bundles. Recognized roots include sklearn, numpy, scipy, pandas, joblib, xgboost, lightgbm, catboost, and statsmodels. For example, a model associated with sklearn causes scikit-learn, numpy, scipy, and joblib to be considered relevant.

Runtime package discovery is intentionally bounded. The environment-capture layer queries installed distribution metadata for a tracked set consisting of scikit-learn, numpy, scipy, pandas, xgboost, lightgbm, catboost, joblib, and statsmodels. Package versions are obtained through Python distribution metadata rather than by importing the corresponding ML packages. Packages that are not installed are omitted. This captured package map is broader than the per-artifact comparison set.

The environment.packages field records installed versions from the tracked package set, while relevant_packages specifies which of those recorded packages participate in model-specific package-drift comparison. Candidate relevant packages absent from the installed-package map are removed before relevant_packages is written to the manifest. This capture-then-filter design reduces unrelated package noise during comparison, but $R(A)$ remains a rule-based approximation rather than a dynamically inferred dependency graph. Its coverage is bounded by the recognized model/component structures and predefined module-to-package bundles.

### *E. Artifact Integrity and Optional Authentication*

After serialization, Modelstamp computes SHA-256 over the staged artifact. Hashing is performed incrementally in 1 MiB chunks, avoiding the need to read the complete artifact into memory. The resulting digest and artifact size are stored in the manifest.

During artifact verification, size is checked before SHA-256 is recomputed. A size discrepancy or digest mismatch produces an artifact-integrity failure. Successful verification therefore establishes artifact consistency relative to the evidence recorded in the manifest, subject to the trust assumptions in Section III.

Optional authentication uses HMAC-SHA256. When a signing key is supplied during persistence, Modelstamp computes an HMAC over a deterministic JSON representation of the manifest evidence. The representation uses sorted keys and compact separators to make the authenticated bytes stable. The signature object records the hmac-sha256 algorithm, authentication digest, and optionally a key_id.

When key_id is used, the identifier is included in the authenticated representation together with the algorithm. Verification can use either a directly supplied shared key or a mapping from trusted key identifiers to keys. A manifest referencing an unregistered key identifier is rejected. Supplying a key registry also requires a signed manifest, preventing an authenticated workflow from silently accepting replacement unsigned evidence.

As established in Section III, this mechanism authenticates evidence under a shared-secret model. Any legitimate holder of the corresponding secret can generate new valid evidence, and HMAC does not independently provide freshness, public publisher identity, or non-repudiation.

### *F. Environment Comparison*

Environment comparison is performed by capturing the current runtime and comparing it with the recorded manifest. The persistence-time manifest may contain versions for the full tracked package set, but package-level comparison projects that state through relevant_packages. For each relevant package, the recorded version is compared with the currently observed version. A changed version or the disappearance of a previously recorded relevant package is represented as a package change.

Modelstamp additionally compares Python version, Python implementation, and platform independently of relevant_packages. Differences in these fields are included as runtime changes in the mismatch report. A mismatch report can therefore contain package changes, runtime changes, an artifact-integrity error, or combinations of these conditions.

This distinction mirrors the formal environment representation in Section II-E: bounded package capture produces $E_s^{cap}$, relevance filtering defines the package projection used by $D_{pkg}$, and runtime-level comparison contributes independently to the broader predicate $D$.

### *G. Verification and Deserialization Ordering*

Modelstamp exposes several operations with different purposes: verify(), check(), inspect(), and load(). inspect() reads and validates the sidecar manifest but deliberately performs neither artifact verification nor deserialization. verify() performs cryptographic verification without deserializing the artifact: it reads and validates the manifest, verifies HMAC when applicable, confirms artifact existence, and verifies artifact size and SHA-256.

load() extends this path with environment comparison and deserialization. Its security-relevant ordering is: read and validate the manifest; authenticate it when applicable; resolve the serialization backend; confirm artifact existence; verify size and SHA-256; compare the environment; and only then deserialize. Environment mismatch handling occurs before deserialization unless the caller explicitly configures mismatches to be ignored.

check() provides a non-deserializing diagnostic path that returns a mismatch report rather than loading the model. It authenticates the manifest before using manifest-controlled environment information, performs environment comparison, and verifies artifact existence, size, and digest. Artifact-integrity failures are captured in the report rather than immediately propagated as an integrity exception.

The distinction between verify() and check() is intentional: verify() focuses on artifact/authentication verification, whereas check() combines integrity and runtime diagnostics without reconstructing the persisted object.

### *H. Mismatch and Failure Behavior*

Environment differences during load() are controlled through the on_mismatch policy. Three behaviors are supported: warn emits an environment-mismatch warning and proceeds to deserialization; raise produces an environment-mismatch exception before deserialization; and ignore suppresses mismatch handling and proceeds after integrity verification. The default behavior is warn.

Integrity and authentication failures are treated differently from environment drift. A size mismatch, SHA-256 mismatch, invalid HMAC, missing required authentication, or untrusted key identifier prevents the normal load() path from reaching deserialization. Missing or malformed manifests produce a manifest error, while a missing artifact after the manifest has been read produces an integrity failure.

This separation reflects the conceptual model established in Sections II and III: environment drift is evidence of changed represented runtime state and can be handled according to caller policy, whereas failed artifact or authentication verification prevents normal deserialization.

Overall, the design binds artifact-integrity evidence and bounded runtime information to a persisted model while applying model-relevance filtering at package-comparison time, without modifying the underlying serialization format. The next section evaluates whether the relevance mechanism, trust-boundary behavior, and verification overhead exhibit the properties defined by the research questions.

## V. EVALUATION METHODOLOGY

### *A. Evaluation Objectives*

The evaluation addresses the three research questions introduced in Section I. RQ1 examines whether Modelstamp distinguishes changes in model-relevant runtime dependencies from unchanged or unrelated environmental changes. RQ2 examines whether artifact-integrity and HMAC-authentication behavior conforms to the trust boundaries defined in Section III. RQ3 characterizes the computational cost of verification as artifact size increases.

The RQ1 and RQ2 experiments are deterministic validation scenarios rather than statistically sampled observations of production failures or attacks. Each scenario specifies a controlled state change and an expected outcome derived from the Modelstamp design and threat model. Agreement between expected and observed behavior therefore demonstrates implementation conformance for the exercised conditions; the scenario counts are not interpreted as estimates of real-world detection rates or security coverage.

The experiments correspond to Modelstamp v0.1.4. For reproducibility, the evaluated repository state is anchored to commit 36fb3284216db327967cc76184eef3a4dedc6c75.

### *B. RQ1: Environment-Drift Evaluation*

RQ1 is evaluated through a CI-enforced matrix of 14 controlled save/check scenarios. Each scenario creates two isolated Python 3.11 environments. A fitted model is persisted under environment E_s, after which modelstamp.check() examines the artifact under environment E_l without deserializing it. Dependency versions are explicitly pinned, and each CI job fails if the set of packages reported as changed differs from the expected set.

The matrix exercises relevant framework changes, supporting-dependency changes, unchanged environments, unrelated-package controls, and broader environmental noise. Framework scenarios include separate scikit-learn patch- and minor-version changes, together with one controlled cross-version change each for XGBoost, LightGBM, and CatBoost. Additional cases exercise scikit-learn drift beneath a LightGBM sklearn-compatible estimator and changes to NumPy, joblib, and SciPy. Negative controls include identical environments and isolated pandas or requests changes. Two broader-noise cases change six unrelated packages, with one of those cases additionally introducing relevant scikit-learn drift.

The broader noise control changes requests, click, rich, pyyaml, packaging, and attrs while keeping the model-relevant scikit-learn stack unchanged. Its companion scenario introduces the same unrelated changes together with a scikit-learn change from 1.5.1 to 1.5.2. This pair tests whether relevance filtering suppresses unrelated environmental variation while retaining a relevant change occurring in the same environment.

For artifact A, let R(A) be the model-relevant package set defined in Section II-E. Restating the package-level component of (10) for the controlled RQ1 comparison, the operational outcome is

$$D_pkg(E_s, E_l; A) = \{p_i \in R(A) \mid v_i \neq v'_i \text{ or } p_i \text{ is absent from the current relevant state}\}. \quad (14)$$

The primary outcome is the set of relevant packages reported as changed rather than merely a binary drift indicator. For a relevant-dependency scenario, the observed set must equal the explicitly expected package set. For unchanged and unrelated-package controls, the expected package-drift set is empty:

$$D_pkg(E_s, E_l; A) = \emptyset. \quad (15)$$

All other pinned dependencies in a scenario are held constant except where the scenario deliberately defines broader environmental noise. Thus, RQ1 tests both sensitivity to selected relevant changes and suppression of selected irrelevant changes. The executable environments are defined by the drift-validation workflow, while the benchmark case script performs the save/check operations.

**TABLE II**
**RQ1 CONTROLLED ENVIRONMENT-DRIFT DESIGN**

| Scenario class | Controlled change | Expecte d drift |
|---|---|---|
| Framework drift | sklearn patch/minor; XGBoost; LightGBM; CatBoost | Yes |
| Wrapper dependency | sklearn under LightGBM sklearn-compatible estimator | Yes |
| Supporting dependency | NumPy; joblib; SciPy | Yes |
| Identical environment | No relevant change | No |
| Unrelated package | pandas-only; requests-only | No |
| Environmenta l noise | Six unrelated packages changed | No |
| Noise + relevant drift | Same noise + sklearn drift | Yes |

### *C. RQ2: Trust-Boundary Evaluation*

RQ2 uses an executable eight-scenario trust-boundary matrix implemented with simple Python objects rather than a machine-learning framework. This isolates Modelstamp's integrity and authentication mechanisms from framework-specific model behavior.

The scenarios comprise: artifact modification with an unchanged manifest; complete replacement of an artifact and unsigned manifest; editing an unsigned manifest's hash while leaving the artifact unchanged; editing unauthenticated model-identity metadata; replacing a signed pair with an unsigned pair; replacing a pair using an untrusted signing key; replacement by a holder of the trusted shared key; and replay of an older but previously valid signed pair. Expected outcomes are derived from the guarantees and non-guarantees defined in Section III.

For candidate artifact A' and recorded digest d, artifact verification evaluates

$$H(A') ?= d. \quad (16)$$

When authentication is enabled, authenticated manifest evidence M is additionally evaluated using

$$HMAC_K(M) ?= t, \quad (17)$$

where K is the trusted shared secret and t is the recorded authentication tag. Modification of an artifact without corresponding valid evidence is expected to produce a digest mismatch. Editing the digest claim itself while leaving the artifact unchanged is likewise expected to fail because the recalculated digest no longer equals the modified claim.

The matrix deliberately contains both expected rejections and expected acceptances. Replacing a signed pair with an unsigned pair must be rejected when HMAC verification is required, as must a replacement authenticated using a key other than the verifier's trusted key. Conversely, possession of the trusted shared secret permits creation of new valid HMAC evidence, and replaying an older valid signed pair remains possible because HMAC does not establish freshness. Editing descriptive identity metadata in an unsigned manifest is also outside the authentication guarantee.

Consequently, RQ2 evaluates conformance to stated guarantees rather than a generic attack-detection percentage. An expected acceptance represents confirmation of a documented non-guarantee rather than a failed security test. The executable matrix is maintained in examples/trust_boundary_matrix.py, with corresponding regression tests in tests/test_trust_boundaries.py.

**TABLE III**
**RQ2 TRUST-BOUNDARY VALIDATION DESIGN**

| Scenario | Property / boundary | Expected |
|---|---|---|
| Artifact modified | Artifact integrity | Reject |
| Unsigned pair replaced | No authentication guarantee | Accept |
| Manifest hash edited | Artifact integrity | Reject |
| Identity metadata edited | Unauthenticated metadata | Accept |
| Signed -> unsigned replacement | Required authentication | Reject |
| Untrusted-key replacement | Key trust | Reject |
| Shared-key forgery | Shared-secret limitation | Accept |
| Valid pair replay | Freshness limitation | Accept |

### *D. RQ3: Verification-Scaling Benchmark*

RQ3 isolates the cost of modelstamp.verify() from model deserialization. The benchmark generates synthetic artifacts of 10 MiB, 100 MiB, and 1024 MiB (1 GiB). Each artifact is accompanied by a manifest containing its byte size and SHA-256 digest but no relevant package entries, allowing the benchmark to concentrate on artifact verification.

Before timing a given size, verify() is invoked once to warm filesystem caches. The benchmark then performs three measured verification runs using time.perf_counter() and reports their median:

$$T_{med} = \mathrm{median}(T_1, T_2, T_3). \quad (18)$$

For artifact size S in MiB, reported verification throughput is

$$\text{Theta} = S / T_{med} \quad \text{MiB/s}. \quad (19)$$

Because verify() performs a streaming SHA-256 pass over the artifact, the benchmark examines whether verification time increases approximately proportionally with artifact size while avoiding deserialization as a confounding operation. The published reference measurements were obtained in a Linux workspace using CPython 3.12 with a warm filesystem cache; CPU, storage characteristics, cold caches, and network filesystems may produce different absolute timings.

### *E. Controls and Outcome Interpretation*

RQ1 uses controls in both directions. The identical-environment case establishes the expected no-drift baseline. The pandas-only, requests-only, and six-package noise scenarios test whether differences outside the selected model-relevant package set produce spurious reports. Relevant framework and supporting-dependency changes test the complementary condition, while the combined noise-plus-scikit-learn scenario determines whether relevant drift remains observable when unrelated changes occur simultaneously.

RQ2 similarly distinguishes claimed properties from explicit non-guarantees. Artifact tampering, altered digest evidence, unsigned replacement under required HMAC verification, and untrusted-key replacement exercise rejection properties. Complete unsigned-pair replacement, unauthenticated descriptive-metadata modification, shared-key-holder forgery, and replay exercise boundaries that hashing or symmetric authentication alone cannot prevent.

Accordingly, 14/14 or 8/8 agreement should not be interpreted as statistical detection accuracy. The matrices are deliberately constructed validation cases whose purpose is to determine whether implementation behavior is consistent with the defined dependency-relevance rules and trust model.

### F. Reproducibility

The RQ1 and RQ2 experiments are executable repository artifacts and are enforced through GitHub Actions. RQ1 constructs isolated save/check environments from pinned package specifications rather than depending on the runner's preinstalled machine-learning environment. RQ2 installs Modelstamp and executes the eight-scenario matrix directly. RQ3 is provided as a standalone benchmark with explicit artifact sizes and repetition count.

The immutable repository commit identified in Section V-A serves as the reproducibility anchor for the evaluated code. The drift workflow and documentation define the 14 controlled environment scenarios, the trust-boundary matrix defines the eight security scenarios, and benchmarks/benchmark_verify.py defines the artifact-size experiment. This separation allows the experimental tables and figures reported in the following section to be traced to executable repository artifacts.

## VI. RESULTS

### A. RQ1: Environment-Drift Results

All 14 controlled scenarios matched their predicted outcomes. Table IV summarizes the environmental change exercised in each scenario and the model-relevant package difference observed by Modelstamp.

**TABLE IV**
**ENVIRONMENT-DRIFT VALIDATION RESULTS**

| Scenario | Controlled change | Observed relevant change |
|---|---|---|
| sklearn patch | 1.5.1 -> 1.5.2 | scikit-learn |
| sklearn minor | 1.5.2 -> 1.6.1 | scikit-learn |
| XGBoost | 2.1.3 -> 3.0.2 | xgboost |
| LightGBM | 4.5.0 -> 4.6.0 | lightgbm |
| LightGBM sklearn wrapper | sklearn 1.3.2 -> 1.9.0 | scikit-learn |
| NumPy relevance | 1.26.4 -> 2.0.2 | numpy |
| joblib relevance | 1.3.2 -> 1.4.2 | joblib |
| pandas noise | 2.2.3 -> 2.3.1 | None |
| Identical environment | Same pinned stack | None |
| SciPy relevance | 1.12.0 -> 1.13.1 | scipy |
| requests noise | 2.31.0 -> 2.32.3 | None |
| Noisy environment | Six unrelated changes | None |
| Noise + relevant drift | Six unrelated + sklearn 1.5.1 -> 1.5.2 | scikit-learn |
| CatBoost | 1.2.7 -> 1.2.8 | catboost |

The results demonstrate two behaviors relevant to RQ1. First, Modelstamp surfaced exact version differences in dependencies selected as relevant to the persisted model. This included framework-level changes and supporting dependencies such as NumPy, SciPy, and joblib. The LightGBM wrapper scenario additionally showed that a scikit-learn change beneath an sklearn-compatible LightGBM estimator remained visible before deserialization.

Second, relevance filtering suppressed the controlled unrelated changes. Neither the pandas-only nor requests-only change produced a reported relevant-package difference. Changing six unrelated packages simultaneously also produced no relevant drift. When the same environmental noise was combined with a scikit-learn change, scikit-learn remained the reported relevant difference.

$$D_pkg(E_s, E_l; A) = \emptyset \quad (20)$$

for the unrelated-noise condition, whereas

$$D_pkg(E_s, E_l; A) = \{\text{scikit-learn}\} \quad (21)$$

when relevant scikit-learn drift was introduced into otherwise equivalent environmental noise. These results do not establish that every dependency selected by Modelstamp is necessary for every model execution, nor that every reported version difference changes model predictions. Within the 14 controlled scenarios, however, the implementation distinguished the exercised model-relevant changes from the exercised unrelated environmental changes according to the expected dependency-selection rules.

*RQ1 finding—Across the 14 controlled scenarios, Modelstamp distinguished the tested model-relevant dependency changes from unchanged and unrelated environmental changes, including a noisy environment in which relevant drift remained independently visible.*

### B. RQ2: Trust-Boundary Results

All eight controlled scenarios matched the behavior predicted by the trust model. Table V reports both the predicted and observed behavior because RQ2 deliberately contains expected rejections and expected acceptances.

**TABLE V**
**TRUST-BOUNDARY VALIDATION RESULTS**

| Scenario | Expected behavior | Observed behavior |
|---|---|---|
| Artifact changed; manifest unchanged | Reject | SHA-256 mismatch |
| Artifact + unsigned manifest replaced | Accept | Accepted |
| Unsigned manifest hash edited | Reject | SHA-256 mismatch |
| Unsigned identity metadata edited | Accept | Accepted |
| Signed pair replaced by unsigned pair | Reject | Manifest not signed |
| Replacement signed with untrusted key | Reject | Invalid signature |
| Replacement signed by shared-key holder | Accept | Accepted |
| Older valid signed pair replayed | Accept | Accepted |

The first and third scenarios confirm the artifact-integrity property. Changing the artifact without updating its recorded digest produced a SHA-256 mismatch. Altering the recorded digest while leaving the artifact unchanged also produced a mismatch because the recomputed artifact digest no longer agreed with the manifest claim.

Authentication strengthened the boundary when a trusted HMAC key was required. Replacing a signed artifact/manifest pair with an unsigned pair was rejected, as was a replacement signed using a different key.

The accepted scenarios expose complementary limits. Without authentication, replacing both the artifact and its unsigned manifest produced a self-consistent pair and verified successfully. Modification of unauthenticated descriptive identity metadata likewise did not affect artifact-integrity verification. A party possessing the trusted shared secret could generate valid authentication evidence for replacement content, and a previously valid signed pair could be replayed because HMAC authenticates content but provides neither freshness nor monotonic version history.

These outcomes distinguish integrity and shared-secret authentication guarantees from provenance and freshness guarantees. The four accepted cases are therefore experimentally demonstrated trust boundaries, not missed detections.

*RQ2 finding—Modelstamp rejected the tested artifact modifications and authentication violations covered by its stated guarantees, while complete unsigned replacement, unauthenticated metadata modification, shared-key-holder replacement, and replay remained possible as explicitly defined limitations.*

### *C. RQ3: Verification-Scaling Results*

Verification time increased with artifact size while measured throughput remained approximately constant across the three benchmark sizes.

TABLE VI
VERIFICATION-SCALING RESULTS

| Artifact size | Median verification time | Throughput |
|---|---|---|
| 10 MiB | 0.032 s | 311.6 MiB/s |
| 100 MiB | 0.326 s | 307.1 MiB/s |
| 1024 MiB | 3.334 s | 307.1 MiB/s |

Increasing artifact size from 10 MiB to 100 MiB increased median verification time from 0.032 s to 0.326 s. At 1024 MiB, median verification time was 3.334 s. Throughput remained within approximately 307–312 MiB/s across the tested range.

$$\Theta = S / T_med \quad (22)$$

This gives 311.6 MiB/s for the 10-MiB artifact and 307.1 MiB/s for both the 100-MiB and 1-GiB artifacts. The near-constant throughput is consistent with the expected cost of streaming SHA-256 verification: the artifact is read once, so verification time grows approximately with artifact size without requiring the complete artifact to be held in memory.

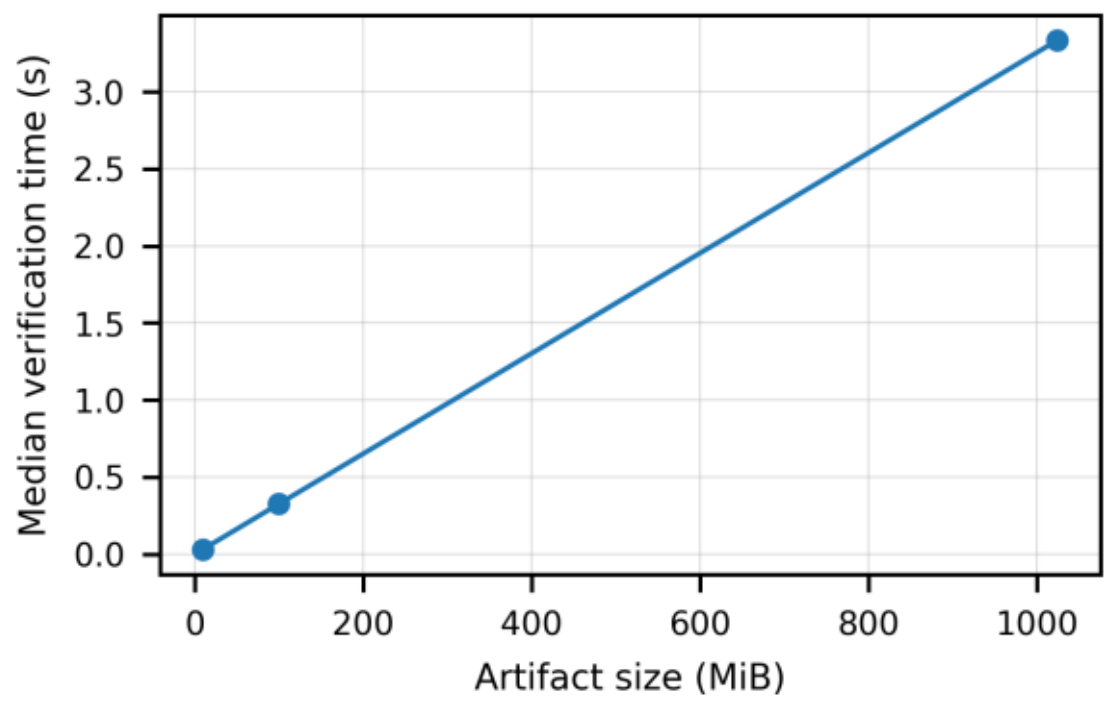

Fig. 2. Median Modelstamp verification time for artifacts of 10 MiB, 100 MiB, and 1 GiB. Each value is the median of three measured runs after one filesystem-cache warm-up. Lines connect measured sizes for visual guidance and do not represent measurements at intermediate artifact sizes.

*RQ3 finding—In the benchmark environment, median verification time increased from 0.032 s for a 10-MiB artifact to 3.334 s for a 1-GiB artifact, while measured throughput remained approximately 307–312 MiB/s, consistent with size-proportional streaming verification.*

### *D. Summary of Findings*

Taken together, the experiments characterize Modelstamp along the three dimensions targeted by the evaluation. The controlled environment-drift matrix showed that the tested relevant dependency changes remained visible while the tested unrelated environmental changes were suppressed. The trust-boundary matrix confirmed both the rejection behavior provided by artifact integrity and optional HMAC authentication and the expected limitations of unsigned evidence, shared-secret authentication, and freshness-free verification. Finally, the artifact-size benchmark showed approximately constant verification throughput across the tested range, with absolute verification time increasing with artifact size.

These findings should be interpreted within the controlled evaluation design described in Section V. The RQ1 and RQ2 scenario matrices validate specified behaviors rather than estimate population-level detection rates, while the RQ3 measurements characterize performance only for the reported benchmark environment.

## VII. RELATED WORK

### *A. Dependency-Aware Persistence and Environment Management*

Dependency recording and cross-version checking already appear in established ML tooling. scikit-learn documents that loading estimators across library versions is unsupported and emits InconsistentVersionWarning when an estimator is unpickled under a different scikit-learn version [8]. PyOD provides a versioned persistence envelope that records PyOD, scikit-learn, NumPy, SciPy, joblib, and Python versions and compares selected dependency versions during load, with warning and strict rejection modes [9]. MLflow similarly packages model dependencies and runtime metadata with saved models and provides environment-management and validation workflows intended to reproduce or test the model's dependency context [10].

These systems establish that dependency metadata and compatibility warnings are not unique to Modelstamp. The distinction investigated here is the combination of bounded persistence-time package capture, an artifact-specific relevance projection for package comparison, cryptographic artifact-consistency evidence, optional shared-secret authentication, and an explicit requirement that the applicable checks occur before Modelstamp's normal deserialization path. Modelstamp therefore should not be read as introducing dependency recording or cross-version mismatch detection; its contribution is the narrower composition and evaluation of these properties within a lightweight reference-state verification layer.

### B. Malicious-Artifact Detection and Safer Deserialization

Unsafe serialization has motivated substantial work on securing persisted machine-learning artifacts. Python serialization mechanisms capable of reconstructing arbitrary objects can expose code-execution behavior during deserialization, making model loading a security-sensitive boundary [2]. Existing defenses consequently include structural analysis of serialized content and mechanisms that restrict operations permitted during deserialization [6], [7].

Policy-based loading constrains which operations or callables may be reconstructed during deserialization [6]. Opcode-based analysis instead characterizes serialized artifacts without executing them and can be used to distinguish malicious from benign artifacts [7]. Modelstamp performs neither function: it does not classify artifact contents as benign or malicious and does not constrain the underlying deserializer. It evaluates artifact consistency relative to recorded integrity evidence and whether represented runtime state has drifted from the reference.

Accordingly, a verified artifact is not necessarily benign or safe to deserialize. A malicious artifact already present when the reference state was established may verify successfully, whereas a benign artifact modified after persistence may fail integrity verification. Malicious-artifact detection, constrained deserialization, and reference-state verification address complementary properties.

### C. Artifact Authentication and Publisher Identity

Cryptographic authentication provides another form of assurance for distributed ML artifacts. Prior work has explored public-key signing infrastructures for binding model artifacts to publisher identities and enabling independently verifiable authentication evidence [3]. Such mechanisms address provenance questions concerning who authenticated or distributed an artifact and can incorporate trust properties beyond artifact-byte consistency.

Modelstamp's optional HMAC mechanism operates under a narrower trust assumption. It authenticates recorded evidence when the producer and verifier share a trusted secret, but possession of that secret also permits generation of new valid authentication evidence. As demonstrated by the trust-boundary evaluation, a shared-key holder can authenticate replacement content, and previously valid authenticated evidence can be replayed because HMAC alone does not establish freshness or monotonic artifact history.

A valid HMAC therefore does not establish public publisher identity or freshness. Modelstamp should not be interpreted as a replacement for public-key publisher authentication, transparency infrastructure, or non-repudiation mechanisms; its HMAC option strengthens artifact/manifest verification within controlled shared-secret workflows.

### D. Dependency Metadata and Broader Environment Capture

Software supply-chain mechanisms commonly represent dependency information through package metadata, lockfiles, or software bills of materials. Prior analysis of Python SBOM generation has shown that reconstructing dependency state from these sources can be incomplete or inaccurate because dependency declarations, package-manager lockfiles, version constraints, and tool implementations vary across the Python ecosystem [4]. Modelstamp avoids this specific reconstruction step by recording concrete installed versions from its bounded tracked set at persistence time, then projecting that captured state through $R(A)$ for package comparison.

Broader reproducibility systems capture substantially more execution context. ReproZip, for example, tracks operating-system calls to collect data dependencies, libraries, and configuration parameters and packages that provenance to support execution in another environment [11]. This represents a stronger and broader reproducibility objective than Modelstamp's selected-state comparison. Modelstamp neither packages a complete execution environment nor attempts to reconstruct one; it deliberately trades completeness for a narrow model-oriented verification signal.

### E. Reproducibility and Artifact Verifiability

Environment recording is related to, but distinct from, reproducibility and broader software-artifact verifiability. Prior research has distinguished reproducible builds from independently verifiable artifacts and examined how dependency and build-environment recovery affects those properties [5]. Empirical work on model reuse likewise identifies provenance, reproducibility, portability, metadata quality, and security as practical concerns [1].

These goals are stronger than the environment property evaluated by Modelstamp. The recorded tracked-package snapshot and relevance-filtered comparison are not intended to reconstruct the complete persistence-time environment, reproduce training, or guarantee identical model outputs. In the notation of Section II-E, $\neg D(E_s, E_l; A)$ means only that no represented package/runtime discrepancy was detected by the defined comparison; it does not imply $E_s \equiv E_l$.

### F. Positioning Modelstamp

The closest adjacent systems demonstrate why Modelstamp's novelty must be stated as a combination rather than as dependency recording alone. Table VII summarizes the resulting design-space position across dependency-aware persistence, lifecycle/environment tooling, malicious-artifact defenses, authentication, supply-chain metadata, and reproducibility mechanisms.

These approaches address related aspects of persisted-model trust, but their guarantees are not interchangeable. Dependency-aware persistence can record or warn about version mismatch [8], [9]; lifecycle tooling can package and validate model dependencies [10]; content-oriented defenses reason about malicious behavior [2], [7]; constrained loaders reason about permitted execution [6]; signing mechanisms establish forms of authentication and provenance [3]; SBOM tooling represents software components [4]; and reproducibility systems capture or reconstruct broader execution context [5], [11].

Modelstamp occupies a narrower position between these mechanisms. It combines artifact-integrity evidence with bounded persistence-time runtime capture, an artifact-specific relevance projection for package comparison, optional shared-secret authentication, and pre-deserialization verification under explicit trust boundaries. It is therefore complementary to, rather than a replacement for, dependency-management systems, malicious-model detection, safe deserialization, public publisher authentication, SBOM generation, or complete reproducibility.

TABLE VII
POSITIONING RELATIVE TO ADJACENT APPROACHES

| Approach class | Primary question | Modelstamp distinction |
|---|---|---|
| scikit-learn guidance | Cross-version estimator compatibility? | Artifact-bound evidence + broader state check |
| PyOD persistence | Do recorded dependencies differ at load? | Adds SHA-256/HMAC + model-relevance filter |
| MLflow environment tooling | What dependencies accompany/reproduce a model? | Narrower local verification; no environment restoration |
| Malicious / constrained loading | Is content malicious or reconstruction permitted? | No maliciousness classification or execution constraint |
| Publisher authentication | Who authenticated/published it? | HMAC does not establish public identity |
| SBOM/dependency tooling | What components/dependencies are represented? | Bounded installed state + relevance-filtered comparison |
| Reproducibility / capture | Can broader computation/environment be reconstructed? | No complete capture or computational reproduction |
| Modelstamp | Do artifact evidence + represented state match reference? | Pre-deserialization artifact + relevance-filtered state verification |

## VIII. DISCUSSION

### A. Interpreting Environment-Drift Verification

The RQ1 results suggest that persisted-model verification benefits from treating artifact state and runtime state as separate but jointly relevant properties. An artifact may remain byte-identical while dependencies associated with its execution change. Artifact hashing alone cannot expose this condition because the artifact digest remains unchanged. Recording selected environment state at persistence time provides an additional reference against which later verification can be performed.

The controlled scenarios also illustrate the importance of dependency relevance. Comparing every installed package would make any environmental modification appear potentially significant, even when the changed package has no identified relationship to the persisted model. In the evaluated scenarios, Modelstamp suppressed changes involving packages treated as unrelated while continuing to surface changes involving the selected model-relevant dependency set. The environmental-noise scenarios are particularly useful in this respect: multiple unrelated changes did not themselves produce relevant drift, while introducing a relevant scikit-learn change into the noisy environment remained detectable.

As formalized in Section II-E, detected environment drift identifies a discrepancy in the represented package or runtime state; it does not establish that model behavior has changed, and the absence of detected drift does not establish equivalence of the complete execution environment. The selected dependency set is an approximation of model relevance. Dependencies may be loaded dynamically, invoked indirectly, supplied through native libraries, or otherwise influence execution without being represented in the recorded set.

Environment verification is consequently best understood as an early warning mechanism. It identifies a condition that may warrant compatibility testing, review, or controlled redeployment rather than determining the semantic effect of the dependency change itself.

### B. Integrity, Authentication, and the Trust Boundary

The RQ2 results illustrate why artifact integrity and artifact authentication must also be distinguished. SHA-256 verification can establish whether candidate artifact bytes match the digest recorded in the manifest. It cannot, by itself, establish whether an attacker has replaced both the artifact and its unsigned reference evidence. A self-consistent replacement pair therefore remains outside the guarantee of hashing alone.

HMAC authentication strengthens this boundary by allowing modification of authenticated evidence to be detected when the attacker does not possess the trusted secret. However, the shared-secret model introduces its own boundary: any party possessing the key can generate valid authentication evidence. The shared-key forgery scenario demonstrates this limitation directly rather than treating key possession as an implicit assumption.

Replay exposes a separate property. A previously valid artifact/manifest pair remains cryptographically valid because neither hashing nor HMAC establishes whether the pair is the most recent authorized state. Preventing replay would require an additional freshness mechanism, such as trusted version state, timestamps anchored to a trusted authority, append-only history, or another mechanism capable of distinguishing current from previously valid evidence.

$$\text{hash match} \Rightarrow \text{artifact consistency with recorded digest}$$

$$\text{valid HMAC} \Rightarrow \text{authentication under possession of the shared key}$$

Neither property independently establishes public publisher identity, authorization history, or freshness. The expected acceptances in the trust-boundary evaluation are therefore as informative as the expected rejections: they identify where additional controls would be required rather than implying that Modelstamp provides a complete artifact-security mechanism.

### *C. Operational Cost of Pre-Deserialization Verification*

RQ3 indicates that the principal verification cost grows approximately with artifact size in the evaluated environment. Throughput remained near 307–312 MiB/s across the three measured sizes, while median verification time increased from 0.032 s at 10 MiB to 3.334 s at 1 GiB.

This behavior is consistent with streaming digest computation, for which artifact bytes must be read to recompute the SHA-256 value. The practical implication is that verification introduces an additional artifact-read operation before deserialization. For small and moderate artifacts, the measured absolute cost was correspondingly small; for larger artifacts, the additional latency becomes more visible.

Whether this cost is acceptable depends on the deployment context. A model loaded once during service initialization has different latency requirements from an artifact repeatedly verified on a request path. Modelstamp therefore does not establish a universally negligible verification overhead. The benchmark instead provides a measured reference from which users can assess the cost relative to their artifact sizes, storage systems, hardware, and loading frequency. The approximately constant measured throughput should likewise not be generalized beyond the benchmark environment.

### *D. Modelstamp as a Complementary Control*

Taken together, the results position Modelstamp as a pre-deserialization reference-state verification mechanism. Its verification path can be summarized as follows:

$$\text{evidence authentication} \rightarrow \text{artifact match} \rightarrow \text{selected environment-drift check} \rightarrow \text{deserialization}$$

As established by the threat-model non-goals in Section III-E and summarized against adjacent mechanisms in Table VII, these checks address only part of the persisted-model trust problem. Modelstamp can therefore be composed with mechanisms for publisher authentication, malicious-artifact detection, constrained deserialization, supply-chain metadata, or deployment-specific compatibility testing when those additional properties are required.

A successful Modelstamp verification should be interpreted narrowly: the artifact satisfies the applicable recorded integrity and authentication checks, and the compared model-relevant environment state satisfies the configured verification conditions. It should not be interpreted as evidence that the artifact is benign, semantically correct, reproducible, or safe to execute.

### *E. Practical Implications*

For model producers, recording integrity evidence and selected environment state at persistence time moves information collection to the point at which the reference artifact and its runtime context are directly available. This avoids relying exclusively on later reconstruction of the persistence environment.

For model consumers, pre-deserialization verification creates an opportunity to identify unexpected artifact or dependency changes before the underlying persistence mechanism loads the model. How a reported environment difference should be handled remains deployment-specific. A strict workflow may reject any relevant drift, while another workflow may treat the report as a trigger for compatibility testing or manual review.

For security-sensitive workflows, the trust-boundary results further indicate that enabling HMAC should not be interpreted as providing publisher provenance. Shared secrets require their own distribution, storage, rotation, and access-control procedures. Where independent publisher verification or non-repudiation is required, a public-key authentication mechanism is more appropriate. These implications reinforce the intended scope of Modelstamp: it makes selected persistence-time assumptions explicit and checkable while leaving compatibility, authorization, execution safety, and deployment policy to mechanisms designed to establish those properties.

## IX. THREATS TO VALIDITY

### *A. Construct Validity*

The evaluation operationalizes environment drift as a difference between the recorded and current versions of dependencies selected by Modelstamp as relevant to a persisted artifact. As formalized in Section II-E, this comparison characterizes differences in the represented environment state rather than semantic changes in model behavior or equivalence of the complete execution environment. Dependency relevance is necessarily an approximation: a package included in the selected set may change version without affecting model behavior, while an execution-relevant dependency may exist outside the identified set.

The RQ2 scenarios similarly evaluate conformance to explicitly defined integrity and authentication properties rather than a general notion of model security. Expected acceptance of unsigned replacement, shared-key-holder replacement, or replay is therefore not treated as a detection failure. These cases exercise properties intentionally outside the corresponding trust guarantees.

### *B. Internal Validity*

The RQ1 and RQ2 matrices consist of deliberately constructed validation scenarios rather than randomly sampled real-world artifacts or attacks. Expected outcomes were derived from the stated design and trust model and then compared with implementation behavior. The resulting 14/14 and 8/8 agreement therefore demonstrates conformance for the exercised cases but must not be interpreted as an estimate of real-world detection accuracy, attack coverage, or failure probability.

Because the scenarios and implementation were developed within the same project, there is also a risk that the evaluation reflects assumptions embedded in the implementation. The inclusion of negative controls, environmental-noise cases, and trust-boundary cases intentionally expected to succeed reduces—but does not eliminate—this risk. Independent evaluation and adversarial test construction would provide stronger evidence.

The dependency-drift experiments additionally use pinned save/check environment pairs. This improves experimental control and reproducibility but samples only specific version transitions. Other version combinations may expose behavior not represented by the current matrix.

### *C. External Validity*

The current dependency-relevance mechanism and RQ1 evaluation are scoped primarily to classical and tabular Python ML frameworks and supporting libraries. The tracked package set currently includes scikit-learn, NumPy, SciPy, pandas, XGBoost, LightGBM, CatBoost, joblib, and statsmodels. Consequently, the reported drift results should be generalized only with caution beyond these ecosystems.

Deep-learning frameworks and formats associated with serialized-model security concerns in prior work [2], [6], [7], including PyTorch-based ecosystems, are not represented in the current tracked dependency set or RQ1 drift matrix. TensorFlow, Transformers, ONNX, and safetensors are likewise outside the evaluated relevance mechanism. The present results therefore do not establish that Modelstamp can correctly identify model-relevant dependency drift for those ecosystems. Extending and evaluating dependency relevance for deep-learning frameworks is left to future work.

The evaluated artifacts and dependency changes also do not represent every model architecture, serialization mechanism, operating system, Python version, hardware configuration, or package interaction encountered in practice. In particular, native libraries, GPU runtimes, system packages, dynamically imported components, and external services may influence execution without being represented in the recorded Python package state.

The trust-boundary evaluation is similarly bounded by its stated adversary model. More complex key-management systems, compromised hosts, malicious persistence-time environments, public-key publisher infrastructures, and distributed transparency mechanisms were not evaluated.

### *D. Performance and Measurement Validity*

The RQ3 benchmark uses three artifact sizes—10 MiB, 100 MiB, and 1 GiB—and reports the median of three measured verification runs for each size after a filesystem-cache warm-up. These measurements characterize the tested benchmark environment rather than a population of deployment environments.

The observed throughput of approximately 307–312 MiB/s should therefore not be interpreted as a universal Modelstamp performance rate. Verification cost can depend on storage hardware, filesystem behavior, operating-system caching, CPU performance, concurrent I/O, artifact location, and other system conditions. Remote or network-backed artifacts may exhibit substantially different behavior.

Similarly, the approximately size-proportional relationship observed across the three measured sizes is consistent with streaming SHA-256 computation but does not constitute a fitted performance model. Intermediate artifact sizes were not measured, and the connecting line in Fig. 2 is provided only for visual guidance.

The benchmark isolates verification from deserialization. This is intentional because RQ3 concerns Modelstamp's verification cost, but it means the reported times should not be interpreted as complete model-loading latency.

### *E. Reproducibility and Evolution*

The evaluation artifacts, scenario definitions, and benchmark procedures are maintained with the public implementation so that the reported behavior can be independently inspected and rerun. Pinned dependency pairs improve reproducibility of the controlled drift experiments, while the benchmark methodology explicitly records the artifact sizes, repetition strategy, and cache treatment used for the reported measurements.

Nevertheless, reproducibility of the experiments does not guarantee identical timing across machines, and future package releases may alter installation compatibility or runtime behavior of the pinned environments. The relevance mapping may also evolve as Modelstamp adds framework support. The results reported here should therefore be interpreted with respect to the evaluated implementation and repository commit identified in Section V-A, rather than assumed to apply unchanged to future versions.

## X. LIMITATIONS AND FUTURE WORK

### *A. Broader Dependency-Relevance Coverage*

As identified in Section IX-C, the current dependency-relevance mechanism and evaluation are concentrated on classical and tabular Python ML frameworks. Extending this mechanism to additional model ecosystems is an important direction for future work. Deep-learning frameworks and associated tooling may introduce different dependency structures, serialization formats, native components, and runtime relationships that are not represented by the current relevance rules.

Future work should therefore evaluate how model-relevant state can be identified for ecosystems such as PyTorch, TensorFlow, and Transformers, as well as formats and runtimes such as ONNX and safetensors. Simply adding more package names to a static mapping may provide incremental coverage but does not address the broader problem of determining which dependencies materially belong to a persisted model's runtime context.

A stronger direction is to investigate more systematic relevance discovery, potentially combining framework-aware rules with artifact metadata, dependency inspection, or configurable user declarations. Any such mechanism should preserve the current objective of avoiding warnings for unrelated environmental changes rather than converging toward indiscriminate recording of the complete installed environment.

### *B. Authentication, Freshness, and Key Management*

The optional HMAC mechanism provides authentication under a shared-secret trust model but does not establish public publisher identity, non-repudiation, or freshness. The trust-boundary evaluation additionally demonstrates that a holder of the trusted secret can authenticate replacement evidence and that previously valid authenticated evidence can be replayed.

Future work could investigate mechanisms for associating verification with trusted version or freshness state and could improve operational support for key identifiers, rotation, and policy enforcement. These extensions should remain separate from the artifact-consistency property provided by hashing.

For workflows requiring public publisher authentication, extending Modelstamp through integration with existing signing infrastructure may be preferable to introducing a bespoke public-key system. Existing work on Sigstore-based model signing [3] provides a natural example of an authentication mechanism that could complement Modelstamp's reference-state verification. Such integration would preserve the separation between Modelstamp's persistence-time evidence and specialized infrastructure for publisher identity and public verification.

### *C. Richer Environment Representation*

Python package versions capture only part of the state capable of influencing model execution. Native libraries, operating-system components, hardware characteristics, accelerator and GPU runtimes, environment variables, and external resources may also affect compatibility or behavior.

Future work could investigate configurable extensions to the recorded environment state for workflows in which these factors are important. For example, GPU-oriented deployments may benefit from recording selected CUDA or accelerator-runtime information, while other environments may require native-library or platform metadata.

The challenge is to increase useful coverage without turning environment verification into a comparison of every observable system property. A richer representation should therefore retain an explicit relevance policy so that additional metadata remains interpretable and does not create excessive environmental noise.

### *D. Integration with Compatibility Validation*

Modelstamp identifies changes in selected recorded state but does not determine whether those changes materially affect model behavior. That distinction should remain part of the design even as the surrounding workflow evolves.

A useful future direction is therefore to expose configurable integration points that allow detected drift to trigger external compatibility validation. A deployment could, for example, invoke an existing model test suite, prediction-regression check, or organization-specific validation process when relevant drift is reported.

Under this design, Modelstamp would surface the condition and provide the recorded/current state needed to make a policy decision; behavioral verification would remain the responsibility of the integrated validation mechanism. This preserves the distinction between state-change detection and behavioral compatibility assessment rather than expanding Modelstamp into a system that claims to establish model correctness or predictive equivalence.

### *E. Broader Empirical Evaluation*

The present evaluation provides controlled evidence for the specified drift, trust-boundary, and performance properties, but broader empirical evaluation would strengthen confidence in how those properties behave across real deployment conditions.

Future studies should include additional model families, framework versions, serialization mechanisms, operating systems, hardware and storage configurations, and larger collections of version transitions. Performance evaluation could similarly use additional artifact sizes, repeated measurements across multiple systems, cold- and warm-cache conditions, and local versus network-backed storage.

Independent evaluation is particularly valuable for the trust model. The current scenario matrices were constructed alongside the implementation and its stated guarantees. Future work should therefore include independently designed adversarial scenarios and third-party evaluation intended to identify assumptions or failure modes not represented by the present test matrices.

Broader evaluation would not change the narrow interpretation of Modelstamp verification, but it would provide stronger evidence about where the mechanism remains reliable, where the relevance rules require extension, and which deployment conditions expose additional trust or performance boundaries.

## XI. CONCLUSION

Persisted machine-learning artifacts are commonly treated as stable objects once serialized, yet their later use depends on software environments that can continue to evolve independently of the artifact bytes. Artifact integrity and environment consistency are therefore distinct verification concerns: an unchanged artifact does not imply an unchanged runtime context, while detected dependency drift does not by itself establish a change in model behavior.

This paper presented Modelstamp, a lightweight Python persistence library that records cryptographic artifact-integrity evidence together with selected model-relevant runtime state and verifies that evidence before deserialization. The controlled evaluation showed that the implementation distinguished the tested relevant dependency changes from unchanged and unrelated environmental changes, including broader environmental-noise conditions. The trust-boundary evaluation confirmed both the intended integrity and authentication behavior and the explicitly stated limitations of unsigned evidence, shared-secret authentication, and replay. The artifact-size benchmark further showed approximately size-proportional verification cost in the evaluated environment.

These findings support a deliberately narrow interpretation of Modelstamp. Successful verification establishes consistency with the applicable recorded artifact evidence and comparison of the selected runtime state; it does not establish that a model is benign, behaviorally correct, reproducible, or safe to deserialize. Those properties require complementary mechanisms.

Modelstamp therefore contributes a pre-deserialization verification layer for making assumptions that often remain implicit in persisted-model reuse—artifact identity and selected runtime context—explicit and checkable. Future extensions can broaden dependency-relevance coverage and integrate additional authentication, environment, and compatibility mechanisms while preserving this separation of responsibilities.